\documentclass[aps,prc,superscriptaddress,showpacs,nofootinbib,eqsecnum]{revtex4}
\usepackage{epsfig,graphics,graphicx}
\usepackage{slashed}
\usepackage{amsmath,lscape,epsfig}
\usepackage{amsfonts}
\usepackage{amsmath}
\usepackage{amssymb}
\usepackage{graphicx,marvosym}

\usepackage{caption}
\usepackage{hyperref,url}

\def\ii{\'{\i}}
\def\beq{\begin{equation}}
\def\eeq{\end{equation}}
\def\beqa{\begin{eqnarray}}
\def\eeqa{\end{eqnarray}}
\def\ban{\begin{eqnarray*}}
\def\ean{\end{eqnarray*}}
\def\bi{\begin{itemize}}
\def\ei{\end{itemize}}

\newcommand*\dashline{\rotatebox[origin=c]{90}{$\dabar@\dabar@\dabar@$}}

\begin{document}

\title{Radiatively Induced Non Linearity in the Walecka Model}

\author{Rafael Cavagnoli} \email{rafael@fsc.ufsc.br }
\affiliation{Departamento de F\'{\i}sica, Universidade Federal de
  Santa Catarina, 88040-900 Florian\'{o}polis, Santa Catarina, Brazil} 
\affiliation{Centro de F\ii sica Te\'orica - Dep. de F\ii sica -
Universidade de Coimbra - P-3004 - 516 - Coimbra - Portugal}

\author{Marcus Benghi Pinto} \email{marcus@fsc.ufsc.br } 
\affiliation{Departamento de F\'{\i}sica, Universidade Federal de
  Santa Catarina, 88040-900 Florian\'{o}polis, Santa Catarina, Brazil} 
\affiliation{Nuclear Science Division,
  Lawrence Berkeley National Laboratory, 94720 Berkeley, CA, USA}

%%%%%%%%%%%%%%%%%%%%%%%%%%%%%%%%%%%%%%%%%%%%%%%%%%%%%%%%%%%%%%%%%%%%%%%%
%%

\begin{abstract}
We evaluate the effective potential for the conventional linear Walecka non 
perturbatively up to one loop. This quantity is then renormalized with a 
prescription which allows   finite vacuum contributions to the three 
as well as four 1PI Green's functions to survive. These terms, which are absent 
in the standard relativistic Hartree approximation, have a logarithmic energy scale dependence 
that can be tuned so as to mimic the effects of $\phi^3$ and $\phi^4$ type of 
terms present in the non linear Walecka model improving quantities such as 
the compressibility modulus and the effective nucleon mass, at saturation, by considering energy
scales which are very close to the nucleon mass at vanishing density.

\end{abstract}


\maketitle

%%%%%%%%%%%%%%%%%%%%%%%%%%%%%%%%%%%%%%%%%%%%%
\section{Introduction}

Quantum hadrodynamics (QHD) is an effective relati\-vistic quantum field theory, 
based on mesons and baryons, which can be used at hadronic energy scales where 
the fundamental theory of strong interactions, quantum chromodynamics (QCD), 
presents a highly nonlinear behavior. The Walecka model \cite{waleckaANP} to be considered 
here represents  QHD by means of a lagrangian density formulated so as to describe 
nucleons interacting through the exchange of an isoscalar vector meson ($\omega$) 
as well as of a scalar-isoscalar meson ($\phi$) which is introduced to simulate 
intermediate range attraction due to the $s$-wave isoscalar pion pairs. 
The original Walecka model (QHD-I) is described by the lagrangian density

\begin{equation}
{\cal L} = {\bar{\psi}} [\gamma_\mu\left(i\partial^{\mu}
- g_v V^\mu)-
(M - g_s \phi) \right ] \psi + \frac{1}{2} (\partial_\mu \phi \partial^\mu \phi- m_s^2 \phi^2)
-\frac {1}{4} F_{\mu \nu}F^{\mu \nu} + \frac{1}{2} m_v^2 V_\mu V^\mu - U(\phi,V) + {\cal L}_{\rm CT}\;, 
\label{lagnlwm}
\end{equation}
where $\psi$, $\phi$ and $\omega$ denote respectively baryon, scalar and vector meson 
fields (with the latter being coupled to a conserved baryonic current). The term $U(\phi,V)$, which describes mesonic self interactions was set to 
zero in the original model so as to minimize the many body effects while the term 
${\cal L}_{\rm CT}$ represents the counterterms needed to 
eliminate any potential ultra violet divergences arising from vacuum computations. 

It is important to recall that, roughly, counterterms are composed by two distinct 
parts the first being  a divergent piece  which exactly eliminates the divergence 
resulting from the evaluation of a Green function at a given order in perturbation 
theory. The second piece is composed by a finite part which is {\it arbitrary} and can 
be fixed by choosing an appropriated renormalization scheme \cite {ramond}. 

The important parameters are the ratios of coupling to masses, $C^2_s$ and $C^2_v$, 
with $C_i^2=(g_iM/m_i)^2$ which are tuned to fit the saturation density 
$\rho_0 = 0.193\, {\rm fm}^{-3}$ and binding energy per nucleon, $BE=-15.75 \, {\rm MeV}$~\cite{waleckaANP}. 
However, the QHD-I predictions for some other relevant static properties of nuclear matter do 
not agree well with the values quoted in the literature. For example, using the Mean 
Field Approximation (MFA) which considers only in medium contributions at the one 
loop level  one obtains that, at saturation, the effective nucleon mass is 
$M^*_{\rm sat} \sim 0.56 \,M$, which is somewhat low, while the compression modulus,  
$K \sim 540\, {\rm MeV}$, is too high. 
In principle, since this is a renormalizable quantum field theory,  vacuum 
contributions (and potential ultra violet divergences) can be properly treated 
yielding meaningful finite results. These contributions  were first considered 
by Chin \cite {chin}, at the one loop level, in the so called Relativistic Hartree 
Approximation (RHA) which produced a more reasonable value for the effective 
mass, $M^*_{\rm sat} \sim 0.72 \, M$. However, the compression modulus remained 
at a  high value, $K \sim 470 \, {\rm MeV}$.  

One could then try to improve the situation by also considering exchange 
contributions since both, MFA and RHA, consider only direct terms in a nonperturbative 
way. When vacuum contributions are neglected this approximation is known as the 
Hartree-Fock (HF) approximation producing
$M^*_{\rm sat} \sim 0.53\, M$ and $K \sim 585 \, {\rm MeV}$ \cite{waleckaANP}.  
By comparing the results from MFA, RHA and HF one sees how vacuum effects can 
improve the values of $M^*_{\rm sat }$ and $K$. Then, the natural question is if the situation 
could be further improved by considering the vacuum in HF type of evaluations. 
The main concern now being the difficulty to deal with overall, nested, and 
overlapping  type of divergences which surely arise due to the self consistent 
procedure.

The complete evaluation of vacuum contributions within direct and exchange terms  
was  performed by Furnstahl, Perry and Serot \cite {furnstahl1} (see Ref. \cite {early} for an early attempt in which only the propagators have been fully renormalized ). This cumbersome 
calculation considers the nonperturbative evaluation and renormalization of the 
energy density up to the two loop level showing  the non convergence of the loop 
expansion. Later, the situation has been addressed with the alternative Optimized 
Perturbation Theory (OPT) which allows for an easier manipulation of divergences \cite {ijmpe}. 
Two loop contributions have been evaluated and renormalized in a perturbative 
fashion with nonperturbative results further generated via a variational criterion. 
However, saturation of nuclear matter could not be achieved with the results 
supporting those of Ref. \cite {furnstahl1}. 

Meanwhile, the compressibility modulus problem has been circumvented by introducing 
some more parameters, in the form of new couplings \cite{bogutabodmer}, to the original Walecka  model. 
One then considers $U(\phi,V)$ appearing in Eq. (\ref{lagnlwm}) as
\begin{equation}
 U(\phi)= \frac{\kappa}{3!}  \phi^3 + \frac{\lambda}{4!} \phi^4 \,,
\label{NL}
\end{equation}
where $\kappa = 2 b M g_s^3$ and $\lambda= 6 c g_s^4$.
This version of QHD is known as the nonlinear Walecka model (NLWM) and the main role of the two additional parameters, $b$ and $c$, is to bring the compression modulus of nuclear matter and the nucleon effective mass under control.
However, one may object to this course of action since the mesonic self interactions will increase the many body effects apart from increasing the parameter space. Notice also that terms like $\phi^n$ ($n \ge 5$) are not allowed since then, in 3+1 dimensions, one would need to introduce coupling parameters with negative mass dimensions spoiling the renormalizability of the original model. apart from increasing the parameter space.

Heide and Rudaz \cite{rudaz} have then realized that it is still possible to keep 
$U(\phi,V)=0$ while improving both $K$ and $M^*_{\rm sat}$. The key ingredient in 
their approach is related to the complete evaluation (regularization and renormalization) 
of divergent vacuum contributions. Regularization is a formal way to isolate the 
divergences associated with a physical quantity for which many different prescriptions 
exist, e.g., sharp cut-off, Pauli-Villars, and Dimensional Regularization (DR). 
Within DR, which was used by Chin, one basically performs the evaluations in 
$d-2\epsilon$ dimensions  taking $\epsilon \to 0$ at the end so that the ultra violet 
divergences show up as powers of $1/\epsilon$.  
However, to keep the dimensionality right when doing $d \to d-2\epsilon$ one has to 
introduce {\it arbitrary} scales with dimensions of energy ($\Lambda$, or the 
related\footnote{The relation between both scales is given by a constant term, 
$\Lambda_{\overline{\rm MS}} = \sqrt{4\pi e^{-\gamma_E}} \,\, \Lambda$, where 
$\gamma_E=-0.5772...$.} $\Lambda_{\overline {\rm MS}}$). Chin has chosen a 
renormalization prescription in which the final results do not depend on the 
arbitrary energy scale while Heide and Rudaz chose one in which such a dependence 
remains, as in most  QCD applications. Since the latter authors also worked at 
the one loop level their approximation became known as the Modified Relativistic 
Hartree Approximation (MRHA) and their main result was to show that it is possible 
to substantially improve $K$ and $M^*_{\rm sat}$ by suitably fixing the energy scale, 
$\Lambda$. Moreover by choosing $\Lambda=M$ the MRHA recovers RHA. In connection with neutron stars, the  MRHA has been applied to the Walecka model in Refs. \cite {cezar}.

Here one of our goals is to treat the Walecka model using a formalism which is closely related 
to the one used in QCD and other modern quantum field theories. Within the QHD model, the   
temperature and density are usually introduced using the real time formalism employed 
in the original work of Walecka. Instead, we use Matsubara's imaginary time 
formalism treating  the divergent integrals  with DR adapted to the modified minimal 
subtraction renormalization
scheme  $\overline {\rm MS}$ \cite{ramond} which constitute the framework most commonly  used within QCD. To obtain the ground state energy density, 
${\cal E}$, we will first evaluate the effective potential (or Landau's free energy), 
$\cal F$, whose minimum gives the pressure, $P$. By choosing appropriate 
renormalization conditions we generate effective three- and four-body couplings, in $\cal F$, which are not 
present at the classical level. As we shall see the numerical values of these effective 
couplings run with the energy scale, $\Lambda_{\overline {\rm MS}}$, allowing for a 
good tuning of $K$ and $M^*_{\rm sat}$  which have their values improved at energy 
scales of about $0.92 \, M$-$0.98 \, M$ ($M =939 \, {\rm MeV}$) while the usual RHA 
results are retrieved for the choice $\Lambda_{\overline {\rm MS}}=M$. 

The MRHA proposed by Heide and Rudaz suggests that if one seeks to minimize 
many-body effects in nuclear matter {\it at saturation},  the choice 
$\Lambda \simeq M^*_{\rm sat}$ is the necessary one. 
Our philosophy is slightly different and perhaps simpler to implement. Since 
possible modifications in the behavior of $K$ and $M^*$ seem to be dictated by the 
presence of $\kappa_{\rm eff} \phi^3$ and $\lambda_{\rm eff}\phi^4$ type of terms 
we shall use the Chin-Walecka renormalization prescription to deal with 
$\phi^n$ ($n=0,1,2$)  vacuum contributions by requiring that their respective 
contributions vanish at zero density (a requirement which was also 
adopted within the MRHA). However, as far as the vacuum contributions related to 
$\phi^3$ and $\phi^4$  are concerned we advocate that one only needs to keep the 
finite energy scale dependent parts in the effective three- and four-body couplings. 
In this way, not only $\kappa_{\rm eff}$ and $\lambda_{\rm eff}$ run with 
$\Lambda_{\overline {\rm MS}}$ but, as we shall see, one also retrieves the RHA 
at $\Lambda_{\overline {\rm MS}}=M$.  

Considering the effective potential  at zero density we will choose an appropriate renormalization prescription for this particular model. 
Since our main goal is to improve $K$ and $M_{\rm sat}^*$ by quantically renormalizing 
$\kappa=0 \to \kappa_{\rm eff}(\Lambda_{\overline {\rm MS}})$ and 
$\lambda=0 \to \lambda_{\rm eff}(\Lambda_{\overline {\rm MS}})$ we can keep 
$M,m_s,m_v$ as representing {\it the} vacuum physical masses  for simplicity. 
This choice  means that, at $k_F=0$, all mass parameters ($M$,$m_s$,$m_v$) represent 
the effective vacuum masses and shall not run with $\Lambda_{\overline {\rm MS}}$ as opposed to $\kappa_{\rm eff}$ and $\lambda_{\rm eff}$. In theories such as QCD the running of the couplings is dictated by the $\beta$ function  whose most important contributions come from the so-called leading logs , e.g.  $\ln(\Lambda_{\overline {\rm MS}}/M)$, which naturally arise in DR evaluations. The application of RG equations to the  effective model Walecka model is beyond the scope of our work \footnote{See Ref. \cite {china} for a RG investigation of the Walecka model.}.  
Nevertheless, our renormalization prescription to obtain a scale dependence so as to better control $K$ and $M_{\rm sat}^*$ is inspired by the leading logs role in the $\beta$ function and the
 the renormalization scheme presented here proposes 
that one should preserve only the scale dependent leading  logs which appear in 
the expressions for $\kappa_{\rm eff} \phi^3$ and $\lambda_{\rm eff}\phi^4$. As 
a byproduct, and contrary to the MRHA case, both quantities will display the same 
scale dependence. Here, this  approximation will be called the Logarithmic Hartree 
Approximation (LHA). The numerical results show that the best LHA predictions for 
$K$ and $M^*_{\rm sat}$ are obtained at energy scales which are only about $5 \%$ 
smaller than that of the RHA, that is 
$ \Lambda_{\overline {\rm MS}} \simeq 0.95 \, M$. This is a nice feature since 
the values of the energy scale and that of the highest mass in the spectrum are 
almost the same whereas in the MRHA the optimum scale, set to be close to 
$M_{\rm sat}^*$ is about $35 \%$ smaller than $M$. From the quantitative point of view, the LHA produces
better results than the MRHA as will be shown.

The work is presented as follows. In the next section the one loop free 
energy is evaluated using Matsubara's formalism. The renormalization of 
the vacuum contributions is discussed in Section III and the complete 
renormalized energy density is presented in Section IV. Numerical results 
and discussions appear in Section V where while our conclusions are presented in Section VI. 
For completeness, in the appendix, we discuss a case in which
$m_s$ does not represent the physical mass.

\section{The Free Energy to One Loop}

In quantum field theories the effective potential (or Landau's free energy), 
${\cal F}$, is defined   as the generator of all one particle irreducible 
(1PI) Green's functions with zero external momentum. The standard textbook 
definition (for one field, $\phi$) reads \cite {ramond}
\begin{equation}
 {\cal F}(\phi_ c) = \sum_{n=0}^{\infty} {\tilde \Gamma}^{(n)}(0) \phi_c^n \,\,,
\end{equation}
where  have absorbed non relevant factors of $i$ and $n!$ by defining ${\tilde \Gamma}^{(n)}(0)=(-i)^n \Gamma^{(n)}(0)/n!$ with $\Gamma^{(n)}(0)$ representing the 1PI $n$-point Green's function and $\phi_c$ 
representing the classical (space-time independent) scalar field. In practice, this 
quantity incorporates quantum (or radiative) corrections to the classical 
potential which appears in the original lagrangian density. While the latter 
is always finite the former can diverge due to the evaluation of momentum 
integrals present in the Feynman loops.
One way to obtain this free energy density is to  perform a functional integration 
over the fermionic fields \cite {ramond}. To one loop this leads to
\begin{equation}
 {\cal F}(\phi_ c,V_c) = - \frac{m_v^2}{2} V_{c,\mu} V^\mu_c + \frac{m_s^2}{2} \phi_c^2 +i \int \frac{d^{4} k}{(2 \pi)^4} {\rm tr}  \ln[\gamma^\mu(k_\mu - g_v V_{c,\mu}) - (M - g_s \phi_c) ] \,\,.
\end{equation}
Notice that this free energy density contains the classical potential (zero loop or tree level term) present in the lagrangian density plus a one loop quantum (radiative) correction represented by the third term. 
Working in the rest frame of nuclear matter we assume that the classical fields are time-like $(V_{c,\mu}= \delta_{\mu,0}V_{c,\mu})$. Then, after taking the trace one can write the free energy as
\begin{equation}
 {\cal F}(\phi_ c,V_{c,0}) = - \frac{m_v^2}{2} V_{c,0}^2  + \frac{m_s^2}{2} \phi_c^2 +i \gamma \int \frac{d^{4} k}{(2 \pi)^4}  \ln[-(k_0 - g_v V_{c,0})^2 + {\bf k}^2 + (M - g_s \phi_c)^2 ] \,\,,
\end{equation}
where $\gamma=4(2)$ is the spin-isospin degeneracy for nuclear (neutron) matter. 
To obtain finite density results one may use Matsubara's imaginary time formalism 
with $k_0 \to i(\omega_n - i\mu)$ where $\mu$ represents the chemical potential 
while, for fermions, $\omega_n= (2n+1) \pi T$ ($n=0,1,...)$ is the Matsubara 
frequency with $T$ representing the temperature. Then, the free energy reads
\begin{equation}
 {\cal F}(\phi_ c,V_{c,0}) = - \frac{m_v^2}{2} V_{c,0}^2  + \frac{m_s^2}{2} \phi_ c^2 - \gamma T \sum_n \int_{}^{} \frac{d^{3} {\bf k}}{(2 \pi)^3}  \ln\{[\omega_n  -(\mu -g_v V_{c,0})]^2 + {\bf k}^2 + (M - g_s \phi_c)^2 ] \,\,.
\end{equation}
The Matsubara's sums can be performed using
\begin{equation}
T\sum_{n=-\infty }^{+\infty }\ln [(\omega _{n}-i{ \mu^\prime}
  )^{2}+E^{2}]=E+T\ln \left[ 1+e^{-\left( E+\mu^\prime \right)
    /T}\right] +T\ln \left[ 1+e^{-\left( E-\mu^\prime \right) /T}\right]
,  
\label{sum1}
\end{equation}
where $E^{2}({\bf k})={\bf k}^{2}+(M-g_s \phi_c) ^{2}$ and $\mu^\prime= \mu - g_v V_{c,0}$. Being interested in the $T=0$ case one may take the  zero temperature
limit of Eq.~(\ref{sum1}) which is given by 

\begin{equation}
\lim_{T\rightarrow 0}T\sum_{n=-\infty }^{+\infty }\ln [(\omega
  _{n}-i\mu^\prime )^{2}+E^{2}({\bf k})]=E({\bf k})+\left[ \mu^\prime -E({\bf k})\right] \theta
(\mu^\prime -E({\bf k}))=\mathrm{max}(E({\bf k}),\mu^\prime )\;.  
\label{sum1tzero}
\end{equation}
Then, at $T=0$ and $\mu \ne 0$, the one loop free energy for the Walecka model becomes
\begin{equation}
 {\cal F}(\phi_ c,V_{c,0}) = - \frac{m_v^2}{2} V_{c,0}^2  + \frac{m_s^2}{2} \phi_ c^2  - \gamma  \int
\frac{d^{3} {\bf k}}{(2 \pi)^3} \left[ \mu^\prime -E({\bf k})\right] \theta(\mu^\prime -E({\bf k})) + \Delta
(\phi_c) \,\,,
\label{Tzero}
\end{equation}
where 
\begin{equation}
 \Delta (\phi_c)= - \gamma \int \frac{d^{3} {\bf k}}{(2 \pi)^3} E({\bf k}) \,\,.
\end{equation}
Power counting shows that $\Delta(\phi_c)$ is a divergent quantity while the $\mu$ dependent term of Eq. (\ref {Tzero}) is convergent due to the Heaviside step function.

\section{ The Renormalized Vacuum Correction Term}

In order to renormalize the vacuum correction term one must first isolate the divergences which is formally achieved by regularizing the divergent integral. Here we use DR performing the divergent integrals in $2\omega=3-2\epsilon$ dimensions. Then, in order to introduce the  ${\overline {\rm MS}}$ energy scale, ${\Lambda_{\overline {\rm MS}}}$, commonly used within QCD one redefines the integral measure as
\begin{equation}
 \int \frac{d^{3} {\bf k}}{(2 \pi)^3} \to \left ( \frac{e^{\gamma_E} \Lambda_{{\overline {\rm MS}}}^2}{4\pi} \right )^{\epsilon/2} \int \frac{d^{2 \omega} {\bf k}}{(2 \pi)^{2 \omega}} \,\,\,,
\end{equation}
where $\gamma_E=-0.5772...$ represents the Euler-Mascheroni constant. Note that, with this definition, irrelevant factors of $\gamma_E$ and $4\pi$ are automatically cancelled but the results of Refs. \cite {chin,ijmpe,rudaz} can be readily reproduced by using $\Lambda_{{\overline {\rm MS}}}= \sqrt{4\pi e^{-\gamma_E}}\Lambda$. The integral can then be performed yielding \cite {ramond} 

\begin{equation}
\Delta(\phi_c) = \gamma \frac{(M - g_s \phi_c)^4}{32 \pi^2} \left\{ \frac{1}{\epsilon} + \frac{3}{2}  - 2\ln \left[ \frac{(M - g_s \phi_c)}{\Lambda_{{\overline {\rm MS}}}} \right]  \right\}   \;.
\end{equation}
As one can see, by expanding the the term proportional to $1/\epsilon$, there are 
five potentially divergent contributions ranging from $g^0$ to $g^4$ while all 
terms of order $g^n$ ($n \ge 5$) are convergent. The divergent terms proportional 
to $\Gamma^{(n)} \phi_c^n$ ($n=0,...,4$) are respectively
\begin{equation}
 {\tilde \Gamma}^{(0)} = \gamma \frac{M^4}{32 \pi^2} \left[ \frac{1}{\epsilon} + \frac{3}{2}  - 2\ln \left( \frac{M}{\Lambda_{{\overline {\rm MS}}}} \right)  \right]   \;,
\label{gamma0}
\end{equation}
\begin{equation}
{\tilde \Gamma}^{(1)}\phi_c =- \gamma \frac{g_s \phi_c M^3}{8 \pi^2} \left[ \frac{1}{\epsilon} + 1  - 2\ln \left( \frac{M}{\Lambda_{{\overline {\rm MS}}}} \right)  \right]   \;,
\end{equation}
\begin{equation}
{\tilde \Gamma}^{(2)} \phi_c^2= \gamma \frac{3 (g_s \phi_c)^2 M^2}{16 \pi^2} \left[ \frac{1}{\epsilon} + \frac{1}{3}  - 2\ln \left( \frac{M}{\Lambda_{{\overline {\rm MS}}}} \right)  \right ]  \;,
\label{gamma2}
\end{equation}
\begin{equation}
{\tilde \Gamma}^{(3)} \phi_c^3= -\gamma \frac{(g_s \phi_c)^3 M }{8 \pi^2} \left[ \frac{1}{\epsilon} - \frac{2}{3}  - 2\ln \left( \frac{M}{\Lambda_{{\overline {\rm MS}}}} \right)  \right]   \;,
\label{gamma3}
\end{equation}
and
\begin{equation}
 {\tilde \Gamma}^{(4)} \phi_c^4= \gamma \frac{(g_s \phi_c)^4 }{32 \pi^2} \left[ \frac{1}{\epsilon} - \frac{8}{3}  - 2\ln \left( \frac{M}{\Lambda_{{\overline {\rm MS}}}} \right)  \right]   \;.
\label{gamma4}
\end{equation}
$ $

The counterterms contained in ${\cal L}_{\rm CT}$ needed to render the free energy 
finite are \cite {chin,ijmpe}

\begin{equation}
 {\cal L}_{\rm CT} = \sum_{n=0}^4 \frac{\alpha_n}{n!}  \phi_c^n \,\,,
\end{equation}
where the $\alpha_n$ coefficients have the general form
\begin{equation}
 \alpha_n \sim g^n_s \left[\frac{1}{\epsilon} + f_n(\Lambda_{{\overline {\rm MS}}}) \right] \,\,.
\end{equation}
$ $

\begin{figure}[tbh]
%\vspace{0.5cm} %\epsfysize=5.5cm
%\epsfig{figure=images/lrha.eps,angle=0,width=7cm}
\includegraphics[width=8.6cm,angle=0]{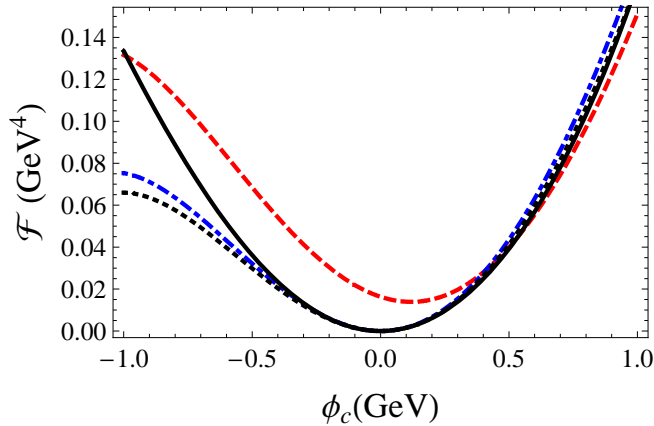}
\caption{(color online) The free energy, in the $\phi_c$ direction, as a function of the classical field for  $\Lambda_{{\overline {\rm MS}}}= 0.9 \, {\rm GeV}$, $M=1 \, {\rm GeV}$, $m_s=0.55\, {\rm GeV}$, and $g_s=1$. The dashed line is the $\overline{\rm MS}$ renormalization scheme result. The dotted-dashed corresponds to $\Gamma^{(0)}=\Gamma^{(1)}=0$ while $\overline{\rm MS}$ is used in the remaining three 1PI function. The same situation but with  $ \Gamma^{(2)}=0$ is represented by the dotted line. The continuous line represents the LHA prescription.}
\label{LRHA}
\end{figure}

Now, within the ${\overline {\rm MS}}$ renormalization scheme generally adopted within 
QCD one simply sets $f_n=0$ and the counterterms have only the bare bones needed to 
eliminate the $1/\epsilon$ poles while the final finite contributions depend on the 
arbitrary energy scale. If one adopts this scheme  within the Walecka model the free 
energy would look like the dashed curve in Fig. \ref {LRHA} which shows $\cal F$ versus $\phi_c$ for the values \footnote {Note that some of the these values are close to the ones which will later  be used in our 
numerical procedure. However, at this stage they are not intended to represent any realistic physical situation apart from letting us  compare possible different shapes of ${\cal F}$.}  $\Lambda_{{\overline {\rm MS}}}= 0.9 \, {\rm GeV}$, $M=1 \, {\rm GeV}$, $m_s=0.55\, {\rm GeV}$, and $g_s=1$. As it is well known, 
within this scheme $M$, $m_s$, and $m_v$ do not represent the measurable physical 
vacuum masses which are instead taken as mass parameters whose values, like the values of the  
couplings, run with $\Lambda_{{\overline {\rm MS}}}$ in a way ultimateley dictated by RG equation. 

\begin{figure}[tbh]
%\vspace{0.5cm} %\epsfysize=5.5cm
%\epsfig{figure=images/diflambdas.eps,angle=0,width=7cm}
\includegraphics[width=8.6cm,angle=0]{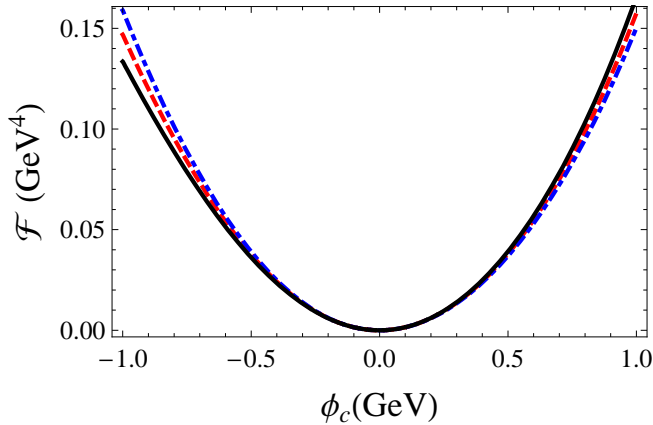}
\caption{(color online) The free energy, in the $\phi_c$ direction, as a function of the classical field for  $\Lambda_{{\overline {\rm MS}}}= 0.9 \, {\rm GeV}$, $M=1 \, {\rm GeV}$, $m_s=0.55\, {\rm GeV}$, and $g_s=1$. All curves represent the LHA prescription at the different scales $\Lambda_{{\overline {\rm MS}}}= 0.9 \, {\rm GeV} < M$ (continuous line), $\Lambda_{{\overline {\rm MS}}}= 1 \, {\rm GeV} = M$ (dashed line) and $\Lambda_{{\overline {\rm MS}}}= 1.1 \, {\rm GeV} > M$. The dashed line also corresponds to the RHA result.}
\label{difLambdas}
\end{figure}

If instead, like Chin, one adopts the so-called on-mass renormalization scheme the counterterms completely eliminate the total contributions 
represented by Eqs (\ref{gamma0}-\ref{gamma4}). Within this choice the results are 
scale independent while $M$, $m_s$, and $m_v$ represent the measurable physical 
masses at zero density whereas the three and four-body mesonic couplings vanish in agreement with the tree level result displayed by the original lagrangian density. 
The free energy generated by this scheme is represented 
by the dashed line in Fig. \ref {difLambdas}. Considering the relevant $K_F=0$ case, let us  
find a hybrid alternative 
scheme between the ${\overline {\rm MS}}$ and the on-mass-shell so that  a 
residual, scale dependent, contribution survives within the three and four 1PI 
Green's function given by Eqs (\ref{gamma3}) and (\ref{gamma4}).

To do that, let us analyze 
each of the arbitrary $f_n$ terms contained in the counterterm coefficients from the physical point of view starting with $f_0$ which is contained in the field independent $\Gamma^{(0)}$. This contribution is 
renormalized by the constant counterterm $\alpha_0$ which can be referred to as the ``cosmological constant'' \cite {lowell}.
In practice, the only effect this term has is to give the zero point energy value and by 
its complete elimination one assures that ${\cal F}(\phi_c=0)=0$ which, 
within the Walecka model, will later assure that the pressure as well as the energy 
density vanish at $k_F=0$. Therefore, as in the on-mass shell prescription, 
we can impose that $f_0$ be exactly equal to the finite part of the 
$\Gamma^{(0)}$ term. It is important to point out that even if one uses the ${\overline {\rm MS}}$ scheme this term can be absorbed in a vacuum expectation value subtraction of the zero point energy so that the exact way in which it done is not too relevant for the present purposes.

The effect of the the linear (tadpole) term $\Gamma^{(1)} \phi_c$ 
is to shift the origin so that the minimum is not at the origin (${\bar \phi}_c \ne 0$) 
as shown by the dashed line of Fig. \ref {LRHA}. 
Also any finite contribution left in the tadpole will cause direct terms to 
contribute to the baryon self energy which, at the present level of approximation, 
means that $M$ does not represent the physical nucleon mass at $k_F=0$, $M^*_{\rm vac}$. 
This can be understood by recalling that the baryon self-energy is $\Sigma_B \sim (g_s/m_s^2){\tilde \Gamma}^{(1)}(\Lambda_{{\overline {\rm MS}}})$ so that the vacuum effective baryon mass is given by  $M^*_{\rm vac} = M+\Sigma_B(\Lambda_{{\overline {\rm MS}}})$ and since $M_{\rm sat}^*=939\, {\rm MeV}$ one sees that $M$, as well as 
$g_s$ and $m_s$, should depend on $\Lambda_{{\overline {\rm MS}}}$. However, for the purposes of controlling $K$ and $M_{\rm sat}^*$   the renormalization of the baryonic vacuum mass from $M$ to $M^*_{\rm vac}$  does not generate the wanted $\phi^3$ and $\phi^4$ vertices. 
Therefore, for simplicity, we can also set $f_1$ so as to completely eliminate the 
tadpole vacuum contribution. This choice for $f_0$ and $f_1$ together with $f_2=f_3=f_4=0$ produces the dot-dashed line of figure 
\ref {LRHA}. The term $\Gamma^{(2)}$ represents a (momentum independent) vacuum correction to the scalar 
meson mass, $m_s$. As in the previous case, getting rid of this term assures that $m_s$ be taken as the 
physical mass simplifying the calculations since $(m_{s, {\rm vac}}^*)^2=m_s^2+ \Sigma_s(\Lambda_{{\overline {\rm MS}}})$ where $\Sigma_s(\Lambda_{{\overline {\rm MS}}}) \sim {\tilde \Gamma}^{(2)}(\Lambda_{{\overline {\rm MS}}})$. Fixing $f_2$ so as to completely eliminate 
the $\Gamma^{(2)}$ contribution produces the dotted line of figure \ref {LRHA}. 
In summary, so far we have adopted the usual Chin-Walecka on-mass shell renormalization 
conditions for $f_1,f_2$, and $f_3$ so that: the vacuum energy is normalized to zero, 
$\phi_c=0$ is the minimum of $\cal F$ (also meaning that $M=M^*_{\rm vac}$), while 
$m_s$ represents the vacuum scalar meson mass. In this approach, none of the vacuum mass parameters present in the 
original lagrangian density run with $\Lambda_{{\overline {\rm MS}}}$. Note that, physically, our choice was also inspired by the NLWM observation that 
the compressibility modulus is improved by the introduction of $\phi^3$ and $\phi^4$ terms which is consistent with our choice of neglecting any corrections to terms proportional to $\phi {\bar \psi} \psi$ and $\phi^2$ which are directly related with the scalar meson and baryon masses. 

Now, taking $f_3=0$ and $f_4=0$ would leave 
us with the wanted $\phi^3$ and $\phi^4$ scale dependent terms. However, inspection 
of Eq. (\ref{gamma3}) and Eq. (\ref {gamma4}) shows that these contributions would 
vanish at different scales, given by $\Lambda_{{\overline {\rm MS}}}=M e^{1/3}$ and 
$\Lambda_{{\overline {\rm MS}}}=M e^{4/3}$ respectively. As already emphasized the NLWM controls the compression modulus with the $\kappa \phi^3$ and $\lambda \phi^4$ 
terms so one can impose that, within our approach,  both  $\kappa_{\rm eff}$ and $\lambda_{\rm eff}$  arise at the same energy scale. This can be  achieved by imposing that 
$ \Gamma^{(3)}=\Gamma^{(4)}=0$ at $\Lambda_{{\overline {\rm MS}}}=M$ in which case the RHA is always reproduced. Finally, in our RG-NLWM inspired prescription we 
also impose that any $\Lambda_{{\overline {\rm MS}}}$ dependence should be left within the leading logs which naturally emerge within DR, as shown by Eqs (\ref{gamma0}-\ref{gamma4}), and which are the main contributing terms to the $\beta$ function.
In this case, the $\alpha_3$ and $\alpha_4$ counterterms also eliminate the 
$\Lambda_{{\overline {\rm MS}}}$ independent constants in Eqs (\ref{gamma3}) and 
(\ref{gamma4}). One then obtains the continuous line of Fig. \ref {LRHA}. The 
complete finite, scale dependent, vacuum contribution is then given by

\begin{eqnarray}
 \Delta_R^{\rm LHA}(\phi_c,\Lambda_{\overline {\rm MS}})&=& -\frac{\gamma}{16 \pi^2} \left[ (M -g_s\phi_c)^4 \, \ln\left( \dfrac{M -g_s\phi_c}{M} \right) + g_s \phi_c M^3 - \frac{7}{2}  (g_s \phi_c)^2  M^2 + \frac{13}{3} (g_s \phi_c)^3 M  - \frac{25}{12} (g_s\phi_c)^4  \right]
\nonumber \\
&+& \frac{\gamma }{4 \pi^2}\left[(g_s \phi_c)^3 M-\frac{1}{4}(g_s \phi_c)^4 \right ] \ln\left( \frac{M}{\Lambda_{{\overline {\rm MS}}}} \right)   \,.
\label{DeltaR}
\end{eqnarray}
$ $

The free energy obtained with this finite vacuum contribution term is shown in Fig \ref {difLambdas} for $\Lambda_{{\overline {\rm MS}}} < M$ (continuous line), $\Lambda_{{\overline {\rm MS}}} > M$ (dot-dashed line) as well as for $\Lambda_{{\overline {\rm MS}}} = M$ (dashed  line) in which case the usual RHA is retrieved. 
As one can check, the first term in Eq. (\ref {DeltaR}) is just the RHA vacuum correction \cite {chin} so that, in view of Eq. (\ref {NL}), one can write
\begin{equation}
\Delta_R^{\rm LHA}(\phi_c,\Lambda_{\overline {\rm MS}})=\Delta_R^{\rm RHA}(\phi_c)+ \frac{\kappa_{\rm eff}}{3!}  \phi_c^3 + \frac{\lambda_{\rm eff}}{4!} \phi_c^4\,\,\,,
\end{equation}
$ $

\noindent where $\kappa_{\rm eff}=2 g_s^3 M b_{\rm eff}(\Lambda_{\overline {\rm MS}})$ and $\lambda_{\rm eff}= 6 g_s^4  c_{\rm eff}(\Lambda_{\overline {\rm MS}})$ with 

\begin{equation}
 b_{\rm eff}(\Lambda_{\overline {\rm MS}})=  \frac{3}{\pi^2} \ln\left( \frac{M}{\Lambda_{{\overline {\rm MS}}}} \right)
\label{kappaeff}
\end{equation}
and $b_{\rm eff}(\Lambda_{\overline {\rm MS}})=-3 \, c_{\rm eff}(\Lambda_{\overline {\rm MS}})$. 
In this way not only $\kappa_{\rm eff}$ and $\lambda_{\rm eff}$ vanish at the same 
scale but an inversion of their respective signs happen at the same time. We have 
then achieved our goal by quanticaly inducing  $\kappa=0 \to \kappa_{\rm eff}(\Lambda_{\overline {\rm MS}})$ 
and $\lambda=0 \to \lambda_{\rm eff}(\Lambda_{\overline {\rm MS}})$ in a way that 
all the scale dependence is contained in the leading logs and also achieving  
$ \kappa_{\rm eff}(\Lambda_{\overline {\rm MS}})=\lambda_{\rm eff}(\Lambda_{\overline {\rm MS}})=0$ 
at $\Lambda_{\overline {\rm MS}}=M$.

For comparison purposes let us quote the MRHA result
\begin{equation}
\Delta_R^{\rm MRHA}(\phi_c,\Lambda)=\Delta_R^{\rm RHA}(\phi_c)+\gamma \frac{(g_s \phi_c)^3}{4 \pi^2}\left[
\ln\left( \frac{M}{\Lambda} \right)-1 + \frac{\Lambda}{M} \right ] -\gamma \frac{(g_s \phi_c)^4}{16 \pi^2}
\ln\left( \frac{M}{\Lambda} \right) \,\,.
\end{equation}
One notices that the major difference between the MRHA and our prescription 
amounts to the finite contribution contained within the cubic term 
where the scale dependence is not restricted to the 
leading log being also contained in  an extra linear term  which does not naturally arise when expands the DR results for the loop integrals in powers of $\epsilon$, as shown by Eqs. (\ref {gamma0}-\ref{gamma4}).

\section{Renormalized Energy Density }

To obtain the thermodynamical potential, $\Omega$, one minimizes the free 
energy (or effective potential) with respect to the fields. That is, 
$\Omega={\cal F}({\bar \sigma}_c, {\bar V}_0) =- P$, where $P$ represents 
the pressure. Then, the LHA renormalized pressure is

\begin{equation}
 P^{\rm LHA} =  \frac{m_v^2}{2} {\overline V}_{c,0}^2  - \frac{m_s^2}{2} {\bar \phi}_c^2  + \gamma \int_0^{k_F}
\frac{d^{3} {\bf k}}{(2 \pi)^3} \left[ (\mu-g_v {\overline V}_{0,c}) -E^*({\bf k}) \right]  - \Delta_R^{\rm LHA} ({\bar \phi}_c,\Lambda_{\overline {\rm MS}})
\label{RenPress}
\end{equation}
$ $

\noindent where $E^*= ({\bf k}^2+M^*)^{1/2}$ with $M^*= M-g_s {\bar \phi}_c$ while the 
Fermi momentum is given by $k_F^2=  (\mu-g_v {\bar V}_{0,c})^2-{M^*}^2$. 
For the vector field one gets 
\begin{equation}
 {\overline V}_{0,c}= \frac {g_v}{m_v^2} \rho_B \,\,,
\end{equation}
$ $

\noindent where $\rho_B= (\gamma k_F^3)/( 6 \pi^2)$ is the baryonic density whereas 
for the scalar field the result is

\begin{equation}
 {\bar \phi}_c = \frac {g_s}{m_s^2}[ \rho_s+\Delta^{\prime \, {\rm LHA}}_R({\bar \phi}_c) ] \,\,\,,
\end{equation}
where 
\begin{equation}
 \rho_s= \gamma \frac{M^*}{2\pi^2}  \int_0^{k_F}d {\bf k} 
\frac{{\bf k}^2 }{E^*({\bf k})}  \;,
\end{equation}
$ $

\noindent represents the scalar density and 

\begin{eqnarray}
 \Delta^{\prime \, {\rm LHA}}_R({\bar \phi}_c)&=&-\frac{\gamma}{4\pi^2}\left[ {M^*}^3 \ln\left( \frac{M^*}{M} \right) + g_s {\bar \phi}_c M^2  - \frac{5}{2}(g_s {\bar \phi}_c)^2 M + \frac{11}{6}(g_s {\bar \phi}_c)^3  \right] 
\nonumber \\
&+&  \frac{\gamma}{4\pi^2}\left[ 3 (g_s {\bar \phi}_c)^2 M -(g_s {\bar \phi}_c)^3
\right ]\ln\left( \frac{M}{\Lambda_{{\overline {\rm MS}}}} \right)     \;.  
\end{eqnarray}
To get the energy density, $\cal E$, one can use the relation 
${\cal E}=-P+\mu \rho_B$ obtaining

\begin{equation}
 {\cal E}^{\rm LHA} =  \frac{g_v^2}{2 m_v^2} \rho_B^2  + \frac{m_s^2}{2} {\bar \phi}_c^2  + \frac{\gamma}{2\pi^2} \int_0^{k_F}
{\bf k}^2 d {\bf k} E^*({\bf k})   + \Delta_R^{\rm LHA} (M^*,\Lambda_{\overline {\rm MS}}) \;,
\label{RenEnerg}
\end{equation}
where

\begin{eqnarray}
 \Delta_R^{\rm LHA}(M^*,\Lambda_{\overline {\rm MS}})&=& -\frac{\gamma}{16 \pi^2} \left[ {M^*}^4 \, \ln\left( \dfrac{M^*}{M} \right) + (M-M^*) M^3 - \frac{7}{2}  (M-M^*)^2  M^2 \right . \nonumber \\
&+& \left . \frac{13}{3} (M-M^*)^3 M  - \frac{25}{12} (M-M^*)^4  \right]
\nonumber \\
&+& \dfrac{\gamma }{4 \pi^2} \left[(M-M^*)^3 M-\frac{1}{4}(M-M^*)^4 \right ]  \ln\left( \dfrac{M}{\Lambda_{{\overline {\rm MS}}}} \right)    \; ,
\label{lha}
\end{eqnarray}
$ $

\noindent and
\begin{equation}
 M^*=M-\gamma \frac{g_s^2}{m_s^2} \frac{M^*}{2\pi^2}  \int_0^{k_F} \frac{{\bf k}^2}{E^*({\bf k})}d {\bf k} - \frac{g_s^2}{m_s^2}\Delta_R^\prime (M^*,\Lambda_{\overline {\rm MS}})  \;,
\label{eff-mass1}
\end{equation}

\noindent where

\begin{eqnarray}
 \Delta^{\prime\, {\rm LHA}}_R(M^*,\Lambda_{\overline {\rm MS}})&=&-\frac{\gamma}{4\pi^2}\left[ {M^*}^3 \ln\left( \frac{M^*}{M} \right) + (M-M^*) M^2  - \frac{5}{2}(M-M^*)^2 M + \frac{11}{6}(M-M^*)^3  \right] 
\nonumber \\
&+&  \frac{\gamma}{4\pi^2}\left[ 3(M-M^*)^2 M -(M-M^*)^3 \right ]  \ln\left( \frac{M}{\Lambda_{{\overline {\rm MS}}}} \right)   \;. 
\label{lhaprime}
\end{eqnarray}

\section{Numerical Results}

Let us now investigate the numerical results furnished by LHA for the baryon 
mass at saturation as well as for the compressibility modulus, with the latter  given by

\begin{equation}
K = \left [ k^2 \frac{\partial^2 }{\partial k^2} \left ( \frac{\varepsilon}{\rho_B} \right ) \right ]_{k=k_F} 
  = 9\left [ \rho_B^2 \frac{\partial^2 }{\partial \rho_B^2} \left ( \frac{\varepsilon}{\rho_B} \right ) \right ]_{\rho_B=\rho_0} \;.
\end{equation}
Table \ref{table-results1} shows the coupling constants and saturation properties 
for some values of the renormalization scale ($\Lambda_{\overline {\rm MS}}$) that 
yield $BE = -15.75$~MeV and $k_F = 1.42 $~fm$^{-1}$~(280.20 MeV). These values are 
chosen just in order to compare with the original Walecka Model (QHD-I) \cite{waleckaANP}. 
The meson masses are $m_s = 512$~MeV and $m_v = 783$~MeV. This table shows 
that some of the best LHA values are obtained with 
$\Lambda_{\overline {\rm MS}}$ values which are very close to $M$. Since at 
$\Lambda_{\overline {\rm MS}}=M$ the RHA result is reproduced one concludes, 
based on our results,  that   slight decrease from the RHA  energy scale 
produces an enormous effect on the values of both, $K$ and $M^*_{\rm sat}$. 

\begin{figure*}[ht]
  \centering
\begin{tabular}{cc}
\includegraphics[width=8.4cm,height=5.7cm,angle=0]{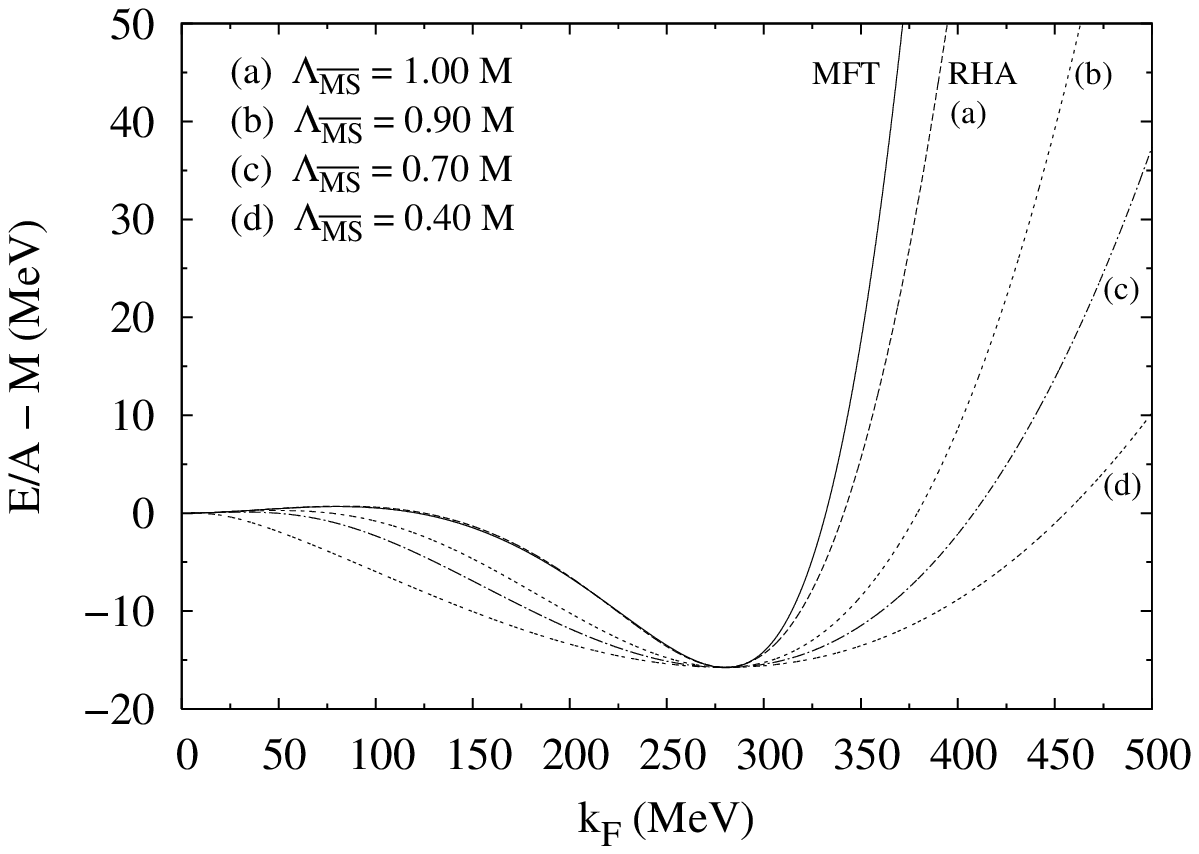} &
\includegraphics[width=8.4cm,height=5.7cm,angle=0]{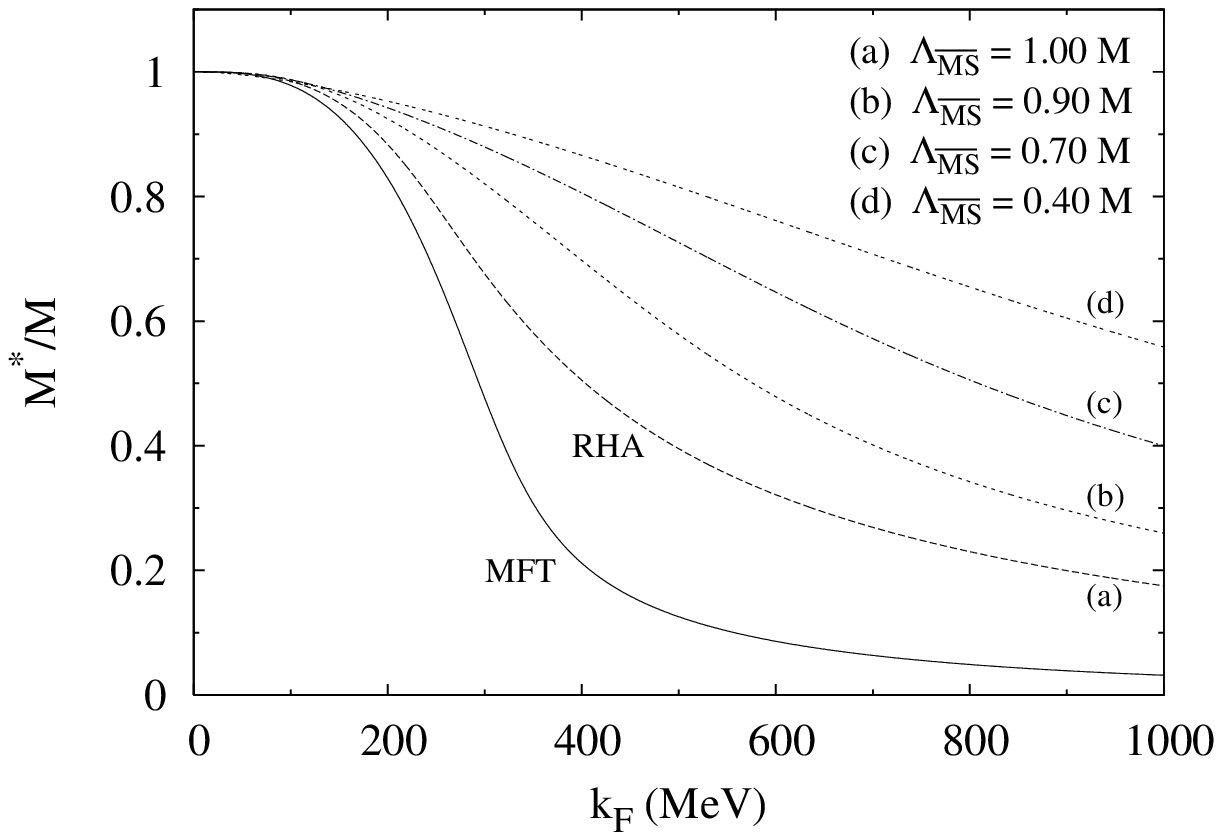}\\
\bfseries (a) &\bfseries (b)\\
\end{tabular}
 \caption{{\bfseries (a)} Binding energy per nucleon as a function of the Fermi momentum for different 
values of our scale $\Lambda_{\overline {\rm MS}}$. We also plot the MFT and RHA results for 
comparison purposes. The saturation properties are: $BE = -15.75$ MeV and 
$k_F = 1.42$~fm$^{-1}$~(280.20 MeV). {\bfseries (b)} Similar as figure (a) but for the effective baryon mass 
$M^*$ as a function of the Fermi momentum for different values of the scale. Note 
that when $\Lambda_{\overline {\rm MS}} = 1$ we reproduce the RHA results.}
 \label{fig1-2}
\end{figure*}

\begin{figure*}[ht]
  \centering
\begin{tabular}{cc}
\includegraphics[width=8.5cm,height=5.7cm,angle=0]{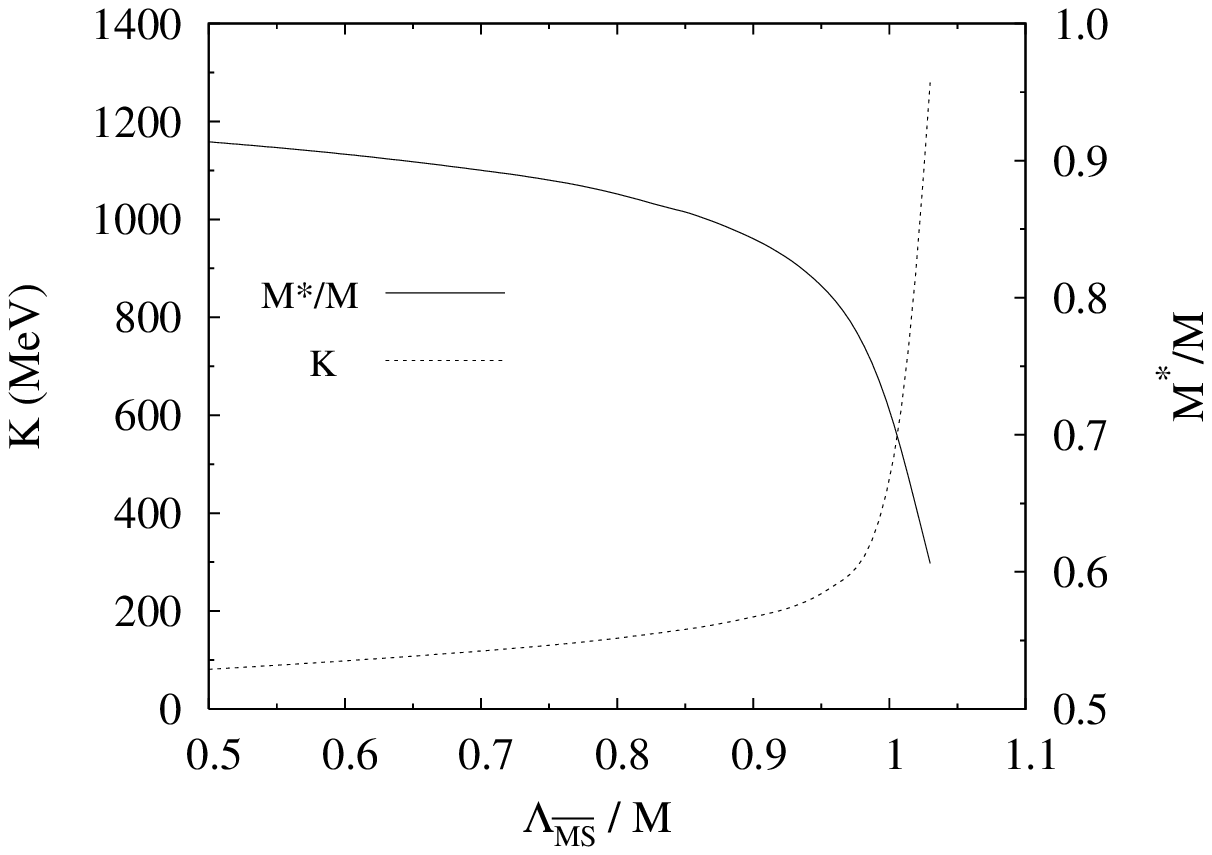} &
\includegraphics[width=8.4cm,height=5.7cm,angle=0]{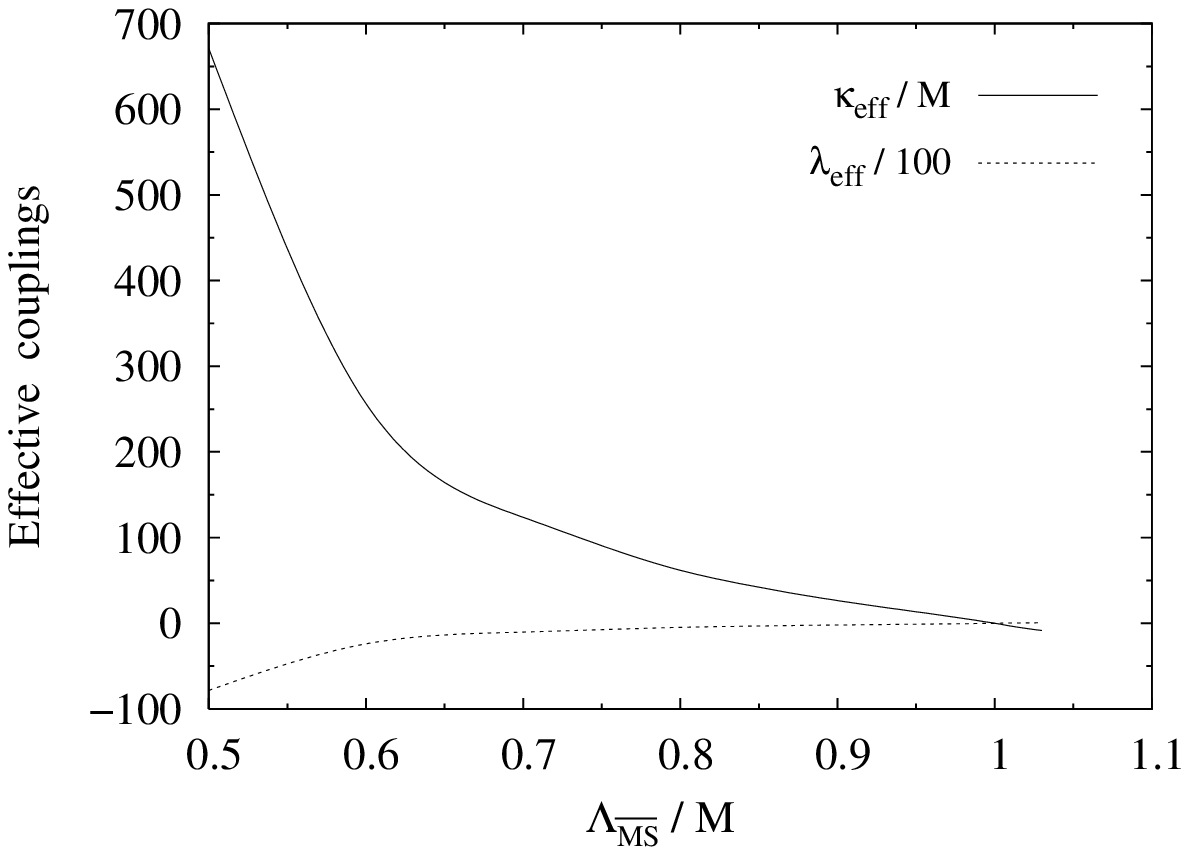}\\
\bfseries (a) &\bfseries (b)\\
\end{tabular}
 \caption{Some properties obtained in table \ref{table-results1} within the LHA 
approach, as a function of the renormalization scale $\Lambda_{\overline {\rm MS}}$ in 
units of the baryon mass. 
{\bfseries (a)} Compression modulus and effective baryon mass $\times$ $\Lambda_{\overline {\rm MS}} /M$ 
and {\bfseries (b)} effective couplings that arise due to the LHA 
as functions of $\Lambda_{\overline {\rm MS}} /M$. They vanish when 
$\Lambda_{\overline {\rm MS}} /M = 1$ and the RHA results are reproduced as one sees 
in table \ref{table-results1}.}
 \label{fig3-4}
\end{figure*}

\begin{table}[th]
\caption{Coupling constants and saturation properties for some values
of the renormalization scale ($\Lambda_{\overline {\rm MS}}$) that yield 
$BE = -15.75$ (MeV) and $k_F = 1.42 $ fm$^{-1}$. The meson masses are 
$m_s = 512$~MeV and $m_v = 783$~MeV.}
\centering
\begin{tabular}{cccccccccc}

\hline
 & $\Lambda_{\overline {\rm MS}}/M$ & $K$ (MeV) & $M^*_{\rm sat}/M$ & $C_v^2$ & $C_s^2$ & $g_v^2$ & $g_s^2$ 
 & $\kappa_{\rm eff}/M$ & $\lambda_{\rm eff}$ \\ 
\hline
 & \hspace{1.5cm} & \hspace{1.5cm} & \hspace{1.5cm} &
\hspace{1.5cm} & \hspace{1.5cm} & \hspace{1.5cm} & \hspace{1.5cm} & \hspace{1.9cm} & \\ 
($\overline {\rm MS}$) & 1.030 & 1279.408 & 0.606 & 171.339 & 176.984 & 119.138 & 52.619 & -6.859 & 49.753 \\ 
($\overline {\rm MS}$) & 1.020 & 910.234 & 0.646 & 151.744 & 184.093 & 105.512 & 54.736 & -4.875 & 36.063 \\ 
($\overline {\rm MS}$) & 1.010 & 639.833 & 0.684 & 132.232 & 185.875 & 91.945 & 55.263 & -2.485 & 18.474 \\ 
($\overline {\rm MS}$) & 1.005 & 542.279 & 0.702 & 123.626 & 185.609 & 85.961 & 55.184 & -1.243 & 9.233 \\ 
%($\overline {\rm M\}$) & 1.000 & 468.140 & 0.718 & 114.740 & 183.300 & 79.782 & 54.497 & 0.000 & 0.000 \\ 
($\overline {\rm MS}$) &\bfseries 1.000 &\bfseries 468.140 &\bfseries 0.718 &\bfseries 114.740 &\bfseries 183.300 &\bfseries 79.782 &\bfseries 54.497 &\bfseries 0.000 &\bfseries 0.000 \\ 
($\overline {\rm MS}$) & 0.990 & 371.437 & 0.745 & 99.784 & 177.933 & 69.383 & 52.901 & 2.351 & -17.099 \\ 
\hline 
($\overline {\rm MS}$) & 0.980 & 314.086 & 0.767 & 88.623 & 173.525 & 61.622 & 51.591 & 4.551 & -32.689 \\ 
($\overline {\rm MS}$) & 0.975 & 294.260 & 0.776 & 84.456 & 172.456 & 58.725 & 51.273 & 5.651 & -40.463 \\
($\overline {\rm MS}$) & 0.970 & 277.989 & 0.784 & 80.307 & 170.683 & 55.840 & 50.746 & 6.694 & -47.684 \\ 
($\overline {\rm MS}$) & 0.960 & 253.249 & 0.798 & 73.691 & 168.648 & 51.240 & 50.141 & 8.811 & -62.391 \\ 
($\overline {\rm MS}$) & 0.950 & 235.660 & 0.809 & 67.923 & 166.440 & 47.229 & 49.484 & 10.855 & -76.356 \\ 
($\overline {\rm MS}$) & 0.940 & 222.493 & 0.818 & 63.025 & 164.576 & 43.824 & 48.930 & 12.875 & -90.058 \\ 
($\overline {\rm MS}$) & 0.920 & 202.507 & 0.832 & 55.411 & 162.702 & 38.529 & 48.373 & 17.054 & -118.611 \\ 
\hline 
($\overline {\rm MS}$) & 0.900 & 188.175 & 0.843 & 49.467 & 161.826 & 34.396 & 48.112 & 21.375 & -148.267 \\ 
($\overline {\rm MS}$) &\bfseries 0.8595 & 166.351 &\bfseries 0.8595 & 40.936 & 164.038 & 28.464 & 48.770 & 31.349 & -218.926 \\ 
($\overline {\rm MS}$) & 0.850 & 162.638 & 0.863 & 39.348 & 165.080 & 27.360 & 49.080 & 33.971 & -237.992 \\ 
($\overline {\rm MS}$) & 0.800 & 144.532 & 0.876 & 32.511 & 172.680 & 22.606 & 51.340 & 49.902 & -357.552 \\ 
($\overline {\rm MS}$) & 0.700 & 118.409 & 0.893 & 23.328 & 200.551 & 16.221 & 59.626 & 99.833 & -770.891 \\ 
($\overline {\rm MS}$) & 0.600 & 98.285 & 0.905 & 17.052 & 256.570 & 11.857 & 76.281 & 206.893 & -1806.98 \\ 
($\overline {\rm MS}$) & 0.500 & 81.019 & 0.914 & 12.239 & 397.387 & 8.510 & 118.147 & 541.142 & -5881.97 \\ 
($\overline {\rm MS}$) & 0.400 & 65.319 & 0.921 & 8.235 & 1290.240 & 5.726 & 383.600 & 4185.070 & -81967.5 \\ 
 &  &  &  &  &  &  &  &  &  \\ 
%(RHA) & - & 468.140 & 0.718 & 114.740 & 183.300 & 79.782 & 62.886 & - & - \\ 
(RHA) &\bfseries - &\bfseries 468.140 &\bfseries 0.718 &\bfseries 114.740 &\bfseries 183.300 &\bfseries 79.782 &\bfseries 54.497 &\bfseries - &\bfseries - \\ 
(MFT) & - & 546.610 & 0.556 & 195.900 & 267.100 & 136.210 & 79.423 & - & - \\ 
\hline
\end{tabular}
\label{table-results1}
\end{table}
Figures \ref{fig1-2} (a) and (b) show the binding energy per baryon, 
$BE = E/A -M$, and the 
effective baryon mass as functions of the Fermi momentum for some  
$\Lambda_{\overline {\rm MS}}$ values,  shown in table 
\ref{table-results1}. One easily sees the effect of considering the vacuum 
contribution and its improvements on the compressibility and the effective mass. As expected, when $\Lambda_{\overline {\rm MS}} = M$, the RHA 
results are recovered. From figures \ref{fig3-4} (a) and (b) it is possible to see some properties 
obtained in table \ref{table-results1} within the LHA 
approach, as functions of $\Lambda_{\overline {\rm MS}}/M$. One notes from figure \ref{fig3-4} (a) that when 
$\Lambda_{\overline {\rm MS}}$ increases the value of the nuclear compressibility ($K$)
also increases and $M^*_{\rm sat}$ decreases. The crossing point 
in figure \ref{fig3-4} (a) represents 
the RHA values of $K$ and $M^*$ which occurs when we set $\Lambda_{\overline {\rm MS}} = M$.
Figure \ref{fig3-4} (b) shows the effective couplings that arise due to the LHA 
as functions of $\Lambda_{\overline {\rm MS}} /M$. Similarly when $\Lambda_{\overline {\rm MS}}$
 reaches the value $M$ the RHA results are recovered and the effective couplings vanish. 


%To compare our numerical results with those provided by the MRHA let us point 
%out that some of most widely accepted values for $M_{\rm sat}^*$ and $K$ are 
%respectively %$0.74 \le M_{\rm sat}^*/M \le 0.85$
%$M_{\rm sat}^*/M \approx 0.74~{\rm to}~0.82$ \cite{johnsonHoren,mahauxSartor,jaminonMahaux}
%and %$200~{\rm MeV} \le K \le 300~{\rm MeV}$ 
%$K \approx 200~{\rm to}~300$~MeV \cite{blaizotK,krivineK,glenK,glenK2,sharmaK}. 

To compare our numerical results with those provided by the MRHA let us make a 
remark concerning the effective nucleon mass. From a non-relativistic analysis of scattering of neutron-Pb nuclei it has 
been found \cite{johnsonHoren} that 
$M_{\rm sat}^*/M \approx 0.74~{\rm to}~0.82$ which can be viewed as  approximately describing 
the Landau effective mass \cite{glenM}. The relativistic 
isoscalar component known as the effective mass defined in Eq. (\ref{eff-mass1}) 
can be called the Dirac effective mass and is related to the Landau effective 
mass. Therefore, the range expected for the Dirac effective mass at saturation 
density lies in the range $M_{\rm sat}^*/M \approx 0.70~{\rm to}~0.80$ whereas
for the nuclear compressibility at saturation the most widely accepted values 
are $K \approx 200\, {\rm MEV}~{\rm to}~300$~MeV \cite{blaizotK}. 
For this range of $K$ and according to table \ref{table-results2} the MRHA predicts 
$M_{\rm sat}^*/M \approx 0.80~{\rm to}~0.85$ for 
$\Lambda/M \approx 1.185~{\rm to}~1.466$. 
However, one should note that this MRHA energy scale 
range is not unique and can also be reproduced with 
$\Lambda/M \approx 0.753~{\rm to}~0.778$
which in turn leads to a rather low range for $M_{\rm sat}^*$ values, %$0.65 \le M_{\rm sat}^*/M \le 0.69$
$M_{\rm sat}^*/M \approx 0.65~{\rm to}~0.69$. Our results, shown in tables \ref{table-results1} 
and \ref{table-results2}, seem to 
produce a better agreement for this range of $K$ giving the {\it unique} 
range $M_{\rm sat}^*/M \approx 0.76~{\rm to}~0.83$
for $\Lambda_{\overline {\rm MS}}/M \approx 0.920~{\rm to}~0.977$ with $\kappa_{\rm eff} >0$ 
and $\lambda_{\rm eff} <0$.
As a last remark we would like to point out that if one chooses $\Lambda_{\overline {\rm MS}}/M = 0.9805$, 
$(g_s/m_s)^2 = 9.468$~fm$^{2}$ and $(g_v/m_v)^2 = 4.879$~fm$^{2}$ the LHA approach 
reproduces the same saturation properties as performed by the so-called GM2 parameter set 
according to \cite{glenGM123}: $K = 300$~MeV, $M_{\rm sat}^*/M = 0.78$, 
$BE = -16.3$~MeV, $k_F = 1.313$~fm$^{-1}$ and $\rho_0 = 0.153$~fm$^{-3}$. The 
resulting effective couplings are: $b_{\rm eff} = 0.005986$ and $c_{\rm eff} = -0.001995$.

%\begin{figure}[ht]
%%\includegraphics[width=8.6cm,height=5.7cm,angle=0]{images/figure1.eps}
%\includegraphics[width=10cm,angle=0]{images/figure1.eps}
%\caption{xxxxx  caption fig 1.} 
%\label{fig1}
%\end{figure}

\begin{table}[th]
\caption{Comparison between the LHA and the MRHA approaches with other estimates.}
\centering
\begin{tabular}{cccc}
\hline
 & $\Lambda_s/M$ & $K$~(MeV) & $M^*/M$ \\ 
\hline
\hspace{2.5cm} & \hspace{2cm} & \hspace{3cm} &  \hspace{1.5cm} \\
MRHA \cite{rudaz} & 1.185 - 1.466 & 300 - 200 & 0.80 - 0.85 \\ 
 &  &  &  \\ 
MRHA \cite{rudaz} & 0.778 - 0.753 & 300 - 200 & 0.65 - 0.69 \\ 
 &  &  &  \\ 
LHA & 0.977 - 0.920 & 300 - 200 & 0.76 - 0.83 \\ 
 &  &  &  \\ 
\hline
Estimates \cite{johnsonHoren,glenM,blaizotK} & - & 300 - 200 & 0.70 - 0.80 \\ 
\hline
% Phenomenology \cite{johnsonHoren,mahauxSartor,jaminonMahaux,blaizotK,krivineK,glenK,glenK2,sharmaK} & - & 300 - 200 & 0.74 - 0.82 \\
\multicolumn{4}{l}{{\footnotesize Where  $\Lambda_s = \Lambda$ for MRHA and LHA is given by: $\Lambda_s = \Lambda_{\overline {\rm MS}}$} .} \\ 

\end{tabular}
\label{table-results2}
\end{table}
In the appendix we show that leaving a leading log 
 dependence also in the two point Green's function, $\Gamma^{(2)}$, only increases the
numerical complexity without producing  results better than the ones generated by the simplest LHA  version 
employed so far.

\section{Conclusions}

We considered the simplest form of the Walecka model to analyze how the values 
of the compressibility modulus as well as the baryon mass, at saturation, can 
be improved by adopting an appropriate renormalization scheme in which cubic 
and quartic effective couplings are radiatively generated. 
With this aim we have evaluated the effective potential to the one loop level 
using Matsubara's formalism to introduce the density dependence. The vacuum 
contributions have been evaluated using dimensional renormalization compatible 
with the ${\overline {\rm MS}}$  renormalization scheme. 

We have then chosen 
the renormalization conditions in such a way so that  all the mass parameters 
appearing in the original lagrangian density represent the physical mass at 
zero density and therefore do not run with the energy scale. 
For our purposes the most important part was to renormalize the values of the cubic 
and quartic terms ($\kappa \phi^3 / 3!$ and $\lambda \phi^4 / 4!$) which vanish at the classical 
(tree) level in the original model. We have then allowed only scale dependent 
logarithms, which naturally arise within DR, to  be 
present in the final finite expressions and, contrary to 
the MRHA prescription, we  obtained that both couplings have exactly the 
same type of scale dependence. In other words, the parameters $b$ 
and $c$ contained in the cubic and quartic terms have been dressed by one 
loop vacuum contributions so that $b=0 \to b_{\rm eff}= 3/\pi^2 \ln (\Lambda_{\overline {\rm MS}}/M)$ 
and $c=0 \to - b_{\rm eff}/3$. 

In our approach each value of the energy scale 
produces only one value for $K$ and $M_{\rm sat}^*$ while two values can be 
obtained within the MRHA. In our case the best values for these physical 
quantities occur at energy scales very close to the highest mass value, $M$. 
 Since the RHA is obtained for $\Lambda_{\overline {\rm MS}}=M$ 
one concludes that a small variation around this value of the energy scale can significantly 
improve both $K$ and $M_{\rm sat}^*$ as shown by our numerical results which 
predict $M_{\rm sat}^*/M \approx 0.76~{\rm to}~0.83$ and $K \approx 200\, {\rm MeV}~{\rm to}~300$~MeV 
at $\Lambda_{\overline {\rm MS}}/M \approx 0.920~{\rm to}~0.977$. These results turn to be 
in excellent agreement with the most quoted estimates $M_{\rm sat}^*/M \approx 0.70~{\rm to}~0.80$ 
and $K \approx 200\, {\rm MeV}~{\rm to}~300$~MeV.  To achieve these $K$ values the 
MRHA predicts either $M_{\rm sat}^*/M \approx 0.80~{\rm to}~0.85$ or  
$M_{\rm sat}^*/M \approx 0.65~{\rm to}~0.69$ 
in the two possible energy scale ranges. Recalling  that at 
$\Lambda_{\overline {\rm MS}}/M=1$ the (RHA) results are $M_{\rm sat}^*/M=0.718$ 
and $K=468.14 \, {\rm MeV}$ one may further appreciate how a small tuning of 
the energy scale within the LHA greatly improves the situation. To compare the LHA with the MRHA we recall that the philosophy 
within the latter is that many-body effects in nuclear matter at saturation can 
be minimized by choosing the energy scale close to $M^*_{\rm sat}$ in which case 
the values $M^*_{\rm sat}/M= 0.731$ and $K=162 \, {\rm MeV}$ are reproduced. 
Although the former seems reasonable the latter seems too low according to 
the above quoted estimates. The philosophy of the LHA, proposed in the present 
work, is to keep only the scale dependent leading logs in the finite parts of the effective 
cubic and quartic couplings.

In practice, the main difference between the two 
approximations is reflected by the fact that the MRHA effective cubic coupling, 
apart from the logarithmic term,  also displays a term which depends linearly 
on the energy scale  accounting for the numerical differences cited above. 
It is worth pointing out that, within  the LHA as well as the MRHA, a given scale sets both $\kappa_{\rm eff}$ and $\lambda_{\rm eff}$ so that both $K$ and $M_{\rm sat}^*$ cannot be separately tuned as in the  NLWM where $\kappa$ and $\lambda$ can be set separately. However, even in an effective theory such as the Walecka model, an increase in the parameter space as the one generated by the NLWM can be viewed as an unwanted feature and the LHA  succeeds in improving the values of $K$ of 
$M_{\rm sat}^*$ without the  drawback of increasing many body effects and parameter space. The method proposed here should be  easy to be implemented  within many 
existing MFA or RHA applications where  $\Delta^{\rm LHA}$ can be added to the energy density (in the MFA case) or used to replace the existing $\Delta^{\rm RHA}$ in a RHA type of calculation.  

In principle the LHA philosophy could be extended to the two loop 
level in a calculation similar to the one performed in Refs. \cite {furnstahl1} and 
\cite {ijmpe}. Then by tuning the energy scale appropriately one could try to 
reduce the size of the two loop corrections producing physically meaningful results.

Our method can be extended to applications related to neutron stars and evaluation of other physical quantities, such as the symmetry energy. 
In particular,  models and/or approximations which lead to low effective masses at saturation 
are not suitable for neutron stars calculations since
as the density increases the effective mass vanishes so quickly that higher densities 
cannot be properly reached as needed \cite{alex-debora}. In principle, the LHA has potential to correct this problem without the need to introduce extra mesonic interactions with their respective parameters.

\begin{acknowledgments}
%\section*{Acknowledgments}
This work was partially supported
Coordenadoria de Aperfei\c{c}oamento de Pessoal de Ensino  Superior
(CAPES, Brazil).  M.B.P.  thanks the Nuclear Theory Group at LBNL, UFSC and
CAPES for the sabbatical leave. We are grateful to D\'{e}bora Menezes, Jean-Lo\"{\i}c Kneur and Rudnei Ramos for comments and suggestions.
\end{acknowledgments}

\appendix*
\section{}
%\section{A test on the $\Gamma^{(2)}$ function}

For completeness, let us check numerically the effects of leaving a leading log 
 dependence also in the two point Green's function with zero external momentum, $\Gamma^{(2)}$, given by Eq. (\ref {gamma2}). Then, 

\begin{equation}
\Delta^{{\rm LHA}}_R(M^*,\Lambda_{\overline {\rm MS}}) \to \Delta^{{\rm LHA}}_R(M^*,\Lambda_{\overline {\rm MS}}) -  (M-M^*)^2 M^2 \frac{\gamma}{16\pi^2}\left[   6 \ln\left( \frac{M}{\Lambda_{{\overline {\rm MS}}}} \right)  \right] \; ,
\end{equation}
and
\begin{eqnarray}
 \Delta^{\prime\, {\rm LHA}}_R(M^*,\Lambda_{\overline {\rm MS}}) \to \Delta^{\prime\, {\rm LHA}}_R(M^*,\Lambda_{\overline {\rm MS}})
-  (M-M^*)M^2 \frac{\gamma}{8\pi^2}\left[  6 \ln\left( \frac{M}{\Lambda_{{\overline {\rm MS}}}} \right)  \right]  \;.  
\end{eqnarray}
In this case, the effective potential gives a first (momentum independent) correction to the effective scalar mass in the vacuum is $m_{s,\rm vac}^*$. Then, for each energy scale, apart from the $BE$ requirement one also has to fix the parameter set so that the effective scalar meson mass, in the vacuum $m_{s, \rm vac}^*= 512 \, {\rm MeV}$. This  effective mass is obtained by considering one loop momentum independent self energy 
\begin{equation}
 (m_{s, \rm vac}^*)^2 = m_s^2 - M^2 g_s^2\frac{\gamma}{8\pi^2}\left[6 \ln\left( \frac{M}{\Lambda_{{\overline {\rm MS}}}} \right)  \right]  \,\,,
\end{equation}
which clearly indicates that $m_s$ (as well as $g_s$) must run with the energy scale. However, this more cumbersome approach has almost no effect in our best results for $K$ and $M_{\rm sat}^*$ as table \ref{table-results-p4}  shows indicating the adequacy of the LHA simple prescription previously adopted.

\begin{table}[th]
\caption{Same as in table \ref{table-results1} for the case in which $\Gamma^{(2)}$ has a non vanishing leading log and $m_s$ runs with the energy scale. }
\centering
\begin{tabular}{ccccccccc}
\hline
 & $\Lambda_{\overline {\rm MS}}/M$ & $K$ (MeV) & $M_{\rm sat}^*/M$ & $C_v^2$ & $C_s^2$ & $g_v^2$ & $g_s^2$ & $m_s$~(MeV) \\ 
\hline  
\hspace{1.2cm} & \hspace{1.3cm} & \hspace{1.5cm} &  \hspace{1.5cm} & \hspace{1.5cm}  & \hspace{1.5cm} & \hspace{1.5cm} & \hspace{1.5cm} & \\
($\overline {\rm MS}$) &\bfseries 1.000 &\bfseries 468.140 &\bfseries 0.718 &\bfseries 114.740 &\bfseries 183.300 &\bfseries 79.782 &\bfseries 54.497 &\bfseries 512.000 \\ 
($\overline {\rm MS}$) & 0.975 & 294.260 & 0.776 & 84.456 & 74.106 & 58.725 & 22.032 & 641.594 \\ 
($\overline {\rm MS}$) & 0.950 & 235.660 & 0.809 & 67.923 & 46.297 & 47.229 & 13.765 & 671.839 \\ 
($\overline {\rm MS}$) & 0.920 & 202.507 & 0.832 & 55.411 & 31.755 & 38.529 & 9.441 & 687.840 \\ 
\hline
\end{tabular}
\label{table-results-p4}
\end{table}

\end{document}